\documentstyle[times,pramana,epsf,psfig,graphicx,rotating,floats]{ias}

\def\beq{\begin{equation}}
\def\eeq{\end{equation}}
\def\bea{\begin{eqnarray}}
\def\eea{\end{eqnarray}}
\def\nnu{\nonumber}
\def\tst{\textstyle}

\def\lbl{\label}
\def\by{\over}

\def\eno#1{Eq.~(\ref{#1})}

\def\al{\alpha}

\def\gam{\gamma}
\def\dta{\delta}
\def\eps{\epsilon}
\def\tta{\theta}

\def\lam{\lambda}

\def\om{\omega}

\def\Dta{\Delta}
\def\Gam{\Gamma}

\def\Om{\Omega}

\def\ptl{\partial}
\def\hf{{1\over2}}
\def\tshf{\tst\hf}
\def\tofro{\leftrightarrow}

\def\apx{\approx}

\def\lp{\left(}
\def\rp{\right)}

\def\ham{{\cal H}}
\def\ket#1{|#1\rangle}

\def\tran#1#2{\langle#1|#2\rangle}

\def\mel#1#2#3{\langle#1|#2|#3\rangle}

\def\bA{{\bf A}}
\def\bB{{\bf B}}
\def\bH{{\bf H}}
\def\bJ{{\bf J}}

\def\xhat{{\bf{\hat x}}}
\def\yhat{{\bf{\hat y}}}
\def\zhat{{\bf{\hat z}}}
\def\nhat{{\bf{\hat n}}}

\def\citn#1#2#3#4#5{#1, #2 {\bf#3}, #4 (#5)} 

\def\jmp{J.\ Math.\ Phys.\ }
\def\npb{Nuc.\ Phys.\ B}

\def\prd{Phys.\ Rev.\ D}
\def\prl{Phys.\ Rev.\ Lett.\ }

\def\Fe8{Fe$_8$}

\def\hsc{{\ham_{\rm sc}}}
\def\Mn12{Mn$_{12}$}

\def\sla{\sqrt{\lam}}
\def\Ecl{E_{\rm cl}}

\def\ntau{\nhat(\tau)}

\def\are{{\cal A}}
\def\ant{\are[\ntau]}
\def\crv{{\cal C}}
\def\surf{{\cal S}}
\def\dedn{{\ptl \Ecl \by \ptl \nhat}}
\def\dndt{{d\nhat \by d\tau}}
\def\rta{\sqrt{1-u_0^2}}

\def\rtl{\sqrt\lam}
\def\rtlb{\sqrt{1 - \lam}}
\def\rtd{\sqrt{(1-u_0^2)(1-\lam)}}
\def\figdir{.}

\begin{document}
\mark{{Spin tunnelling}{Anupam Garg}}
\title{Spin tunnelling in mesoscopic systems}

\author{Anupam Garg}

\address{Department of Physics and Astronomy, Northwestern University,
Evanston, Illinois 60802, USA}



\abstract{
We study spin tunnelling in molecular magnets as an instance of a
mesoscopic phenomenon, with special emphasis on the molecule
\Fe8. We show that the tunnel splitting between various pairs
of Zeeman levels in this molecule oscillates as a function of
applied magnetic field, vanishing completely at special points in
the space of magnetic fields, known as diabolical points. This
phenomena is explained in terms of two approaches, one based on
spin-coherent-state path integrals, and the other on a generalization
of the phase integral (or WKB) method to difference equations.
Explicit formulas for the diabolical points are obtained for
a model Hamiltonian.}

\maketitle

\section{Introduction}
The last five years or so have seen a burst of activity in the physics
and chemistry of molecular magnets. This activity is fuelled to a great
extent by the ability of chemists to synthesize organic molecules
containing magnetic ions such as such as Mn, Fe, and Co, which then
form molecular solids, in many cases of high crystallinity and
homogeneity. More than
5000 different magnetic molecular clusters are known, and about a
100 of these show some behaviour or quirk that is of interest from the
persepctive of physics \cite{cds99,jv99}. These molecules are
intermediate bewteen simple paramagnetic salts such as
CuSO$_4$.K$_2$SO$_4$.6H$_2$O or
Ce$_2$Mg$_3$(NO$_3$)$_{12}$.24H$_2$O, in which the magnetic entites are
single transition metal or rare earth ions, and superparamagnetic
particles of submicrometer size. Thus, they show phenomena such as
hyeteresis at the molecular level \cite{ns95,fs96,tlbd96}, but since
the magnetic entities are molecules and thus well defined with
identical size, shape, and orientation, and since
the interactions of the magnetic and nonmagnetic degrees of freedom are
relatively well understood, there is the potential for understanding
the energy relaxation process in great detail. Some of these issues are
similar to those that arise in the study of magnetization reversal of
small magnetic particles, and it is hoped that the molecular systems will
offer insights into the latter, which has obvious implications for
magnetic recording and storage technologies.

Our purpose in this article is rather different. The molecular
systems have total spin of the order of 10, and magnetocrystalline
anisotropies
of few tens of Kelvin in energy. The quantum mechanical dynamics of the
spin are then profitably viewed in terms of tunnelling in many cases,
and it is of interest to look for such behavior. At present, the
greatest interest is in
the molecule [Fe$_8$O$_2$(OH)$_{12}$(tacn)$_6$]$^{8+}$ (henceforth
abbreviated to \Fe8), which is a molecular complex in which
antiferromagnetic interactions between the Fe$^{3+}$ ions within a
molecule lead to a ground state with a total spin of 10. This molecule
shows clear evidence for spin tunnelling in the form of a
tunnel splitting that oscillates as a
function of a static external magnetic field \cite{ws99}. One way to
understand this effect is in terms of the spin-coherent-state path
inetgral for spin. The kinetic term in this path integral has the
properties of a Berry phase, which can allow for interference of spin
trajectories \cite{ldg92,vdh92}. In fact, using this property, the
oscillations were found theoretically \cite{agepl93}, without
knowing of the existence of \Fe8. However, another way to understand the
effect, which has the advantage of using only classical methods of
analysis, is in terms of a discrete analogue of the phase integral
or WKB method.

In Sec. 2 we shall give a brief discussion of the experiments on \Fe8.
and in Secs. 3 and 4, we shall describe the instanton and discrete
phase integral (DPI) methods for calculating spin tunnel splittings.

\section{Summary of experimental results on \Fe8}

Electron paramagnetic resonance and other measurements show that at
low temperature, the
Fe$^{3+}$ spins in one molecule of \Fe8 behave as a single large spin
of magnitude 10, with an anisotropy energy that is well described by
the Hamiltonian,
\beq
\ham_0 = k_1 J_x^2 + k_2 J_y^2,
     \label{ham0}
\eeq
with $J=10$, $k_1 \apx 0.33\ $K, and $k_2 \apx 0.22\ $K. 
The g-factor of the net spin is very close to 2, with very little
anisotropy in the $g$-tensor.

If we think of \eno{ham0} as a classical energy function for a classical
vector $\bJ$, we see that the energy is a minimum when $\bJ\|\pm \zhat$,
and a maximum when $\bJ\| \pm \xhat$.
We refer to $\xhat$, $\yhat$, and $\zhat$
as the hard, medium, and easy axes respectively. When the problem is
treated quantum mechanically, the quantum states corresponding to the
two degenerate minima at $\pm\zhat$ should be able to mix together via
tunnelling. This picture remains true when we add a magnetic field in
the {\it xy\/} plane. The Hamiltonian is now
\beq
\ham = -k_2 J_z^2 + (k_1 - k_2)J_x^2
        -g \mu_B (J_x H_x + J_y H_y), \label{ham1}
\eeq
where we have subtracted out a constant $k_2\bJ\cdot\bJ$. The minima
are now moved off the $\pm\zhat$ axes toward the equator, but they
continue to be degenerate, and should mix by tunnelling. Indeed, as
$H_\perp$, the magnitude of the field in the {\it xy\/} plane,
increases, both the angle through which the spin must tunnel and the
energy barrier decrease, and we expect that the splitting $\Dta$ will
increase.

At this point, two questions arise. First, how should one think of the
tunnelling of a spin? And second, how large is the splitting?
One way to answer the first question is to 
regard the last three terms as perturbations that give rise
to transitions between various Zeeman levels or eigenstates of $J_z$.
As usual, we denote the $J_z$ value by $m$. The $J_x^2$ term gives rise to
$\Dta m = 2$ transitions, and thus mixes $m=-10$ with $m = +10$ via the
$-8$, $-6$, $\ldots$, $+8$ states. The $H_x$ and $H_y$ terms give rise
to $\Dta m= 1$ transitions, and mix $m=-10$ with $+10$ via all intermediate
levels $-9$ to $+9$ (see Fig.~1a). This picture allows us to think of
spin tunnelling in direct analogy with a particle tunnelling through an
energy barrier. It is further obvious that if we also apply a field along
the $z$ direction so as to tune the energies of the $-10$ and $+9$ states
to resonance, we can also think of tunnelling between these states. More
generally, we can consider tunnelling between a state with $m=m_i$ on the
negative $m$ side and $m = m_f$ on the positive $m$ side.
\begin{figure}
\centerline{\psfig{figure=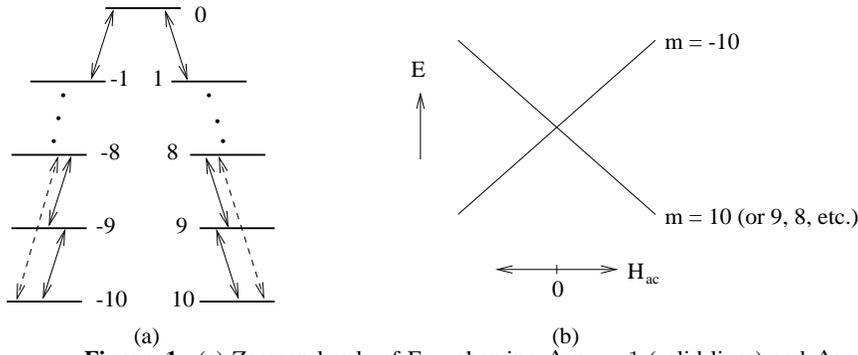,width=0.9\linewidth}}
\caption{(a) Zeeman levels of \Fe8, showing $\Dta m =1$ (solid lines)
and $\Dta m = 2$ (dashed lines) transitions. (b) The Landau-Zener-%
St\"uckelberg process.}
\end{figure}

The second question is in some sense the main subject of this article. Yet
it is useful to have an approximate answer before one embarks on a detailed
calculation. For a massive particle tunnelling in a symmetric double well,
the splitting can quite generally be written in the form
\beq
\Dta = c_1\om_0 \left( {S_0 \by 2\pi } \right)^{1/2} e^{-S_0},
     \label{Dtaapx}
\eeq
where $\om_0$ is the classical small oscillation frequency about the minima,
$S_0$ is the tunnelling action or WKB exponent, and $c_1$ is a constant
of order unity. Further, for a smooth potential, $S_0$ can be written
as $c_2 V_0/\om_0$, where $V_0$ is the energy barrier, and $c_2$ is another
constant of order unity, generally close to 5. For the quartic potential
$V(x) = V_0(x^2-a^2)^2/a^4$, e.g., $c_1 = 4\sqrt3$ and $c_2 = 16/3$.
To apply this approximate formula to the spin problem, we may take
$V_0 = k_2 J^2$, but we still need $\om_0$. This, however, can be found by
noting that the classical vector $\bJ$ can be given dynamics via Hamitlon's
equation and Poisson brackets:
\beq
{d \bJ \over dt} = \{\bJ, \ham\}_{\rm PB}
      = - \bJ \times {\ptl\ham \over \ptl\bJ}. \lbl{eomJ}
\eeq
Then, ${\dot J}_x = 2k_2 J_y J_z$, and ${\dot J}_y = -2k_1 J_x J_z$. 
Linearizing around the equilibrium state $\bJ = J\zhat$, we find that
$\om_0 = 2(k_1k_2)^{1/2} J$. For \Fe8, $\om_0 \apx 5.4\,$K, and if we
use the approximate formula (\ref{Dtaapx}) with the same $c_1$ and $c_2$
as for the quartic double well, we find that $\Dta \apx 60\,$nK.
This is an extraordinarily low splitting, of the order
of 1\,KHz in frequency units. At first sight, the question of how to
experimentally detect such a low splitting would appear to be entirely
moot, since each molecule experiences stray magnetic fields $H'$
of abcond-mat.garg.9112out 200\,Oe
of dipolar and hyperfine origin. Consequently, the Zeeman levels
$m = \pm 10$ are shifted from perfect degeneracy by an amount
$\eps \simeq g\mu_b J H' \simeq 0.1\,$K, which is enormous compared to
$\Dta$. Now to see any resonance at all between two states, the energy
bias between them must be comparable to or less than the assumptive
tunnelling amplitude. In \Fe8, exactly the opposite is true, and it follows
that left to itself, a molecule will almost never tunnel at all.

Wernsdorfer and Sessoli circumvent this problem by applying an ac
magnetic field along $\zhat$, in the shape of a triangular wave of
period $\tau$ and amplitude $H_0$. As $H_z$ changes with time, the energies
of the $m = \pm 10$ states move in opposite directions, and cross at
some point in the cycle. This induces what are known as Landau-Zener-%
St\"uckelberg (LZS) transitions \cite{lzs32,llqm}.
In the limit where the crossing
point is passed rapidly, the $-10 \tofro 10$ transition probability $\gam$ in
each passage is very low, $\propto \Dta^2 /(dH/dt)$. Since the stray
fields are fluctuating randomly,
different passages may be assumed to be independent,
with no phase correlation between successive passages. The net LZS rate,
or transition probability per unit time is given by
\beq
\Gam_{\rm LZS} = {2 \over \tau}\gam
           \apx {\pi \Dta^2 \over (g\mu_B\hbar) \Dta m H_0}
                           \quad (\gam \ll 1).
           \lbl{ratelzs}
\eeq
Wernsdorfer and Sessoli measure this rate by measuring the rate at
which the magnetization, after having being initially saturated along the
$-\zhat$ direction, say, relaxes to its equilibrium direction
under the influence of the
ac longitudinal field, and a dc field, which may or may not have a
longitudinal component.

The outcome of the experiments is shown in Fig.~2. (The temperature at
which these data are taken is 0.36\,K.) Our earlier expectation
that $\Dta$ will increase monotonically as $H_\perp$ increases is seen to
be false.
When $\bH \| \yhat$ ($\phi = 90^{\circ}$), the behavior is monotonic,
but when $\bH \| \xhat$ ($\phi = 0^{\circ}$),
one finds that $\Dta$ oscillates with $H_x$!
\begin{figure}
\centerline{\psfig{figure=\figdir/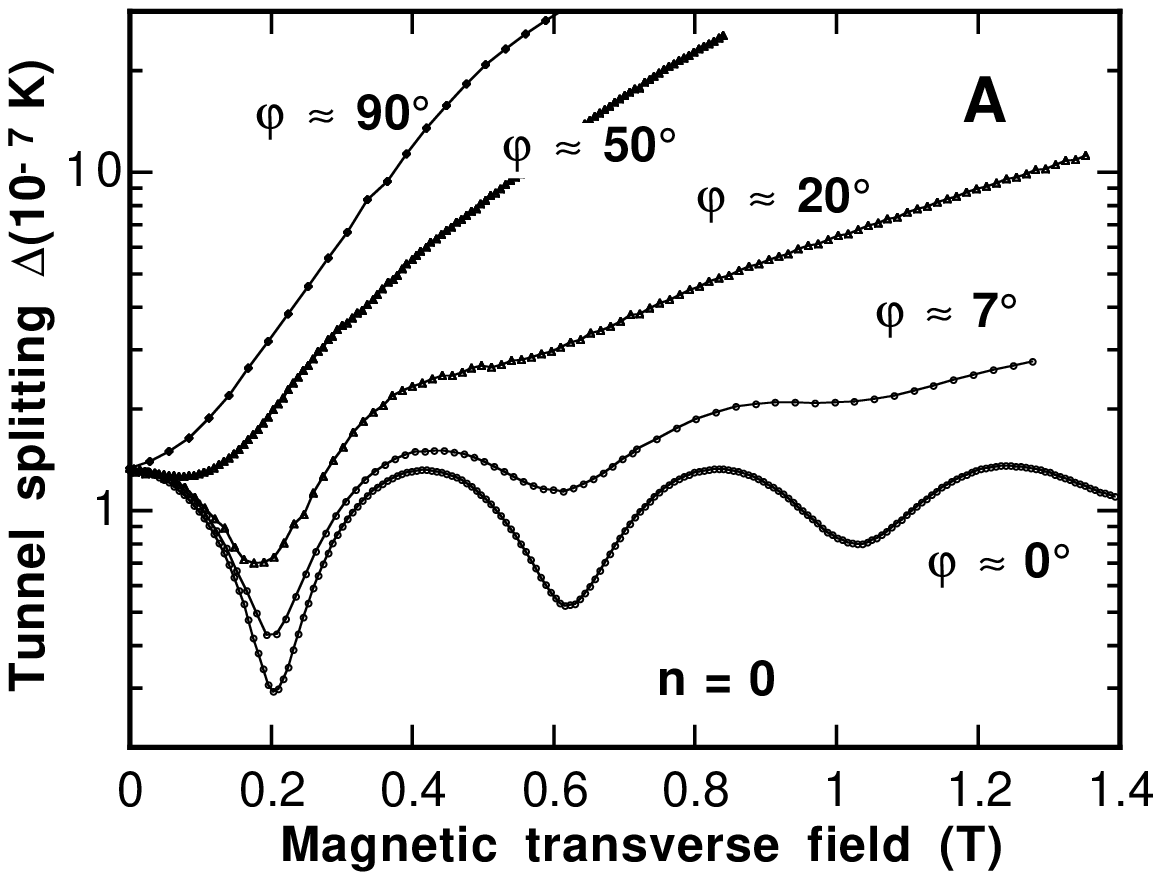,height=5cm,%
width=0.46\linewidth}
\psfig{figure=\figdir/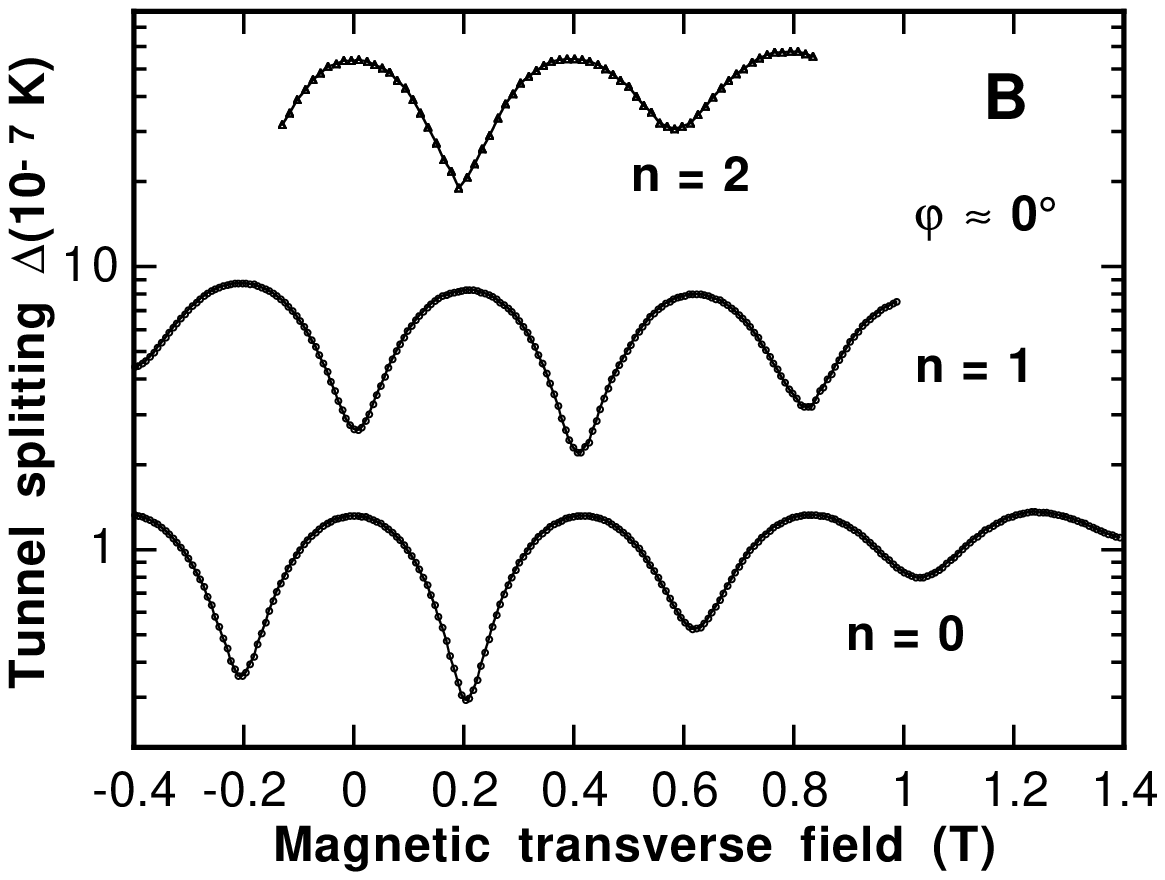,height=5cm,width=0.46\linewidth}}
\caption{Measured splittings [6] for \Fe8 for (a) $-10 \tofro 10$
transitions for various orientations of $\bH$ in the {\it xy\/} plane,
and (b) for $\bH\|\xhat$ between the states $m =-10$ and $m= 10 -n$.
Figure courtesy of Dr. Wernsdorfer.}
\end{figure}

\section{Instanton approach to spin tunnelling}
\subsection{General comments on instantons}
For massive particles, the instanton approach is an elegant and effective
way to calculate tunnel splittings, especially between ground
states \cite{jl67,sc77+}. Extended to spin, the basic idea is to
examine the imaginary time transition amplitude
\beq
U_{21} = \mel{\nhat_2}{e^{-\ham T}}{\nhat_1},
      \lbl{U21}
\eeq
where $\nhat_{1,2}$ are the minima of the classical energy
\beq
E_{\rm cl}(\nhat) = \mel{\nhat}{\ham}{\nhat},
\eeq
and $\ket{\nhat}$ is a maximal-projection spin coherent state,
i.e., an eigen state of $\bJ\cdot\nhat$ with eigenvalue $J$.

In the $T \to \infty$ limit, only the lowest two energy eigenstates
contribute in a spectral decomposition of \eno{U21}, and since these differ
in energy by $\Dta$, we get
\beq
U_{21} \sim \sinh(\Dta T), \label{Uasymp}
\eeq
ignoring inessential prefactors. If we can calculate $U_{21}$ in this limit,
comparison with \eno{Uasymp} will yield $\Dta$. This calculation is done
by appealing to the spin-coherent-state path integral
\beq
U_{21} = \int_{\nhat_1}^{\nhat_2} [d\nhat]\, e^{-S[\ntau]}.
  \label{Upathint}
\eeq
The paths $\ntau$ all run from $\nhat_1$ at $-T/2$ to $\nhat_2$ at
$T/2$, and $S[\ntau]$ is the imaginary time or Euclidean action for the path.
(This is why the exponent is $-S$ rather than $iS/\hbar$.)

Since $T$ is tending to $\infty$, we can evaluate \eno{Upathint} via
the steepest descents approximation. The dominant paths, known as instantons,
are just those that minimize the action, i.e. they are solutions
to the classical equations of motion. The simplest such paths consist of
a single transit from $\nhat_1$ to $\nhat_2$. If the scale over which $\Ecl$
varies is $V$, then it follows from \eno{eomJ} that the time scale for
this transit is $\tau_0 \sim J/V$. For the Hamiltonian
(\ref{ham0}), e.g., this
time scale is $\om_0^{-1} \sim J^{-1} (k_1 k_2)^{-1/2} \ll T$. Hence,
the spin spends most of its time near the end points $\nhat_{1,2}$, and the
actual transit takes place in a very short time interval. (Hence the name
instanton.) In the same way, we can find anti-instantons, solutions that
go from $\nhat_2$ to $\nhat_1$. We now note that \eno{eomJ} is autonomous,
i.e., does not depend on $\tau$ explicitly. Therefore a translation of the
center of the instanton yields an equally good classical
path. Secondly, one can fit an arbitrary number of instantons followed by
antiinstantons into the interval $T$. One finds that the $n$-instanton
contribution to $U_{21}$ is proportional to $T^n$, and the full series is
that of a $\sinh$ \cite{sc77+}. The $\Dta$ which is obtained in this way
can be written as
\beq
\Dta = D \exp(-S_{\rm inst}),
\eeq
where $S_{\rm inst}$ is the action for a {\it single instanton} path, and
$D$ is a prefactor arising from doing the path-integral over small fluctuations
about the instanton trajectory. If more than one instanton exists, we must
add together the corresponding contributions from all of them.

In the application of this formalism to any tunnelling problem, it turns
out that the calculation of $S_{\rm inst}$ is often very easy, while the
calculation of $D$ is quite hard. (Indeed, it turns out that the spin analogue
of the van Vleck determinant for the general massive particle propagator
has only recently been found in a completely logically consistent way
\cite{hgs87,eak95,vvps95,spg00}.) Since the most interesting aspects of
the \Fe8 problem can be understood at the level of the instanton action
alone, and since the absolute scale of $\Dta$ can be found from the DPI
method, we shall forego the calculation of $D$ in this article.

\subsection{Action for spin}
The interesting features of the spin problem lie in the form of the action,
\beq
S[\ntau] = iJ \ant + \int_{-T/2}^{T/2} \Ecl[\ntau] d\tau.
  \lbl{Sofn}
\eeq
The term $\ant$ is the kinetic term, and has the mathematical structure of
a Berry phase. We write it as
\beq
\ant = \int_{\surf}d\Om
       \qquad (\crv = \ptl\surf = \ntau - \nhat_R), \lbl{Asurf}
\eeq
by which we mean that $\ant$ is the area of the patch $\surf$ on the unit
sphere, whose boundary $\crv$ is the closed curve
formed by the path $\ntau$ and a reference path $\nhat_R$ (taken backwards)
running from $\nhat_1$ to $\nhat_2$ (Fig.~3). The reference path $\nhat_R$
is arbitrary, but must be the same for all $\ntau$ in the path integral.
Its choice is equivalent to fixing the gauge. However, since we have
defined $\are$ as a geometrical quantity, an area, it does not depend on how
we choose coordinates on the unit sphere, and
it is obviously nonsingular. Note also that on the sphere, each closed
curve divides the sphere into two areas, and one cannot say which is inside
or outside. (This is easily seen by thinking of the seam on a tennis ball.)
The two areas differ by $4\pi$, however, so $e^{-iJ\are}$ is identical for
the two of them for either integral or half-integral $J$, and hence this
ambiguity does not affect the path integral.
\begin{figure}
\centerline{\psfig{file=\figdir/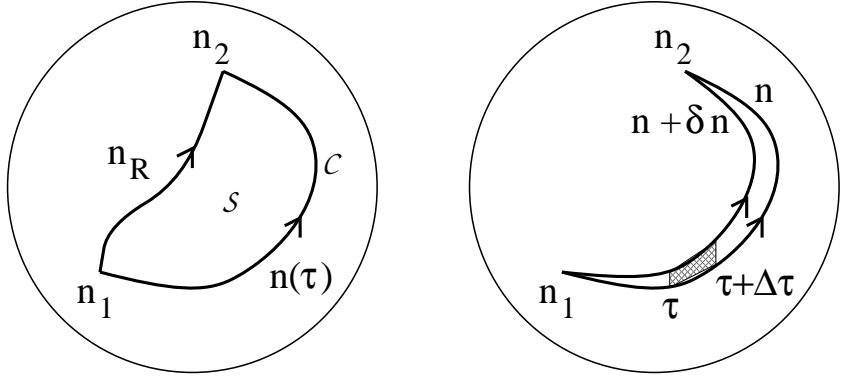,height=5cm}}
\caption{The kinetic term in the spin action, and its variation.}
\end{figure}

Instead of deriving \eno{Sofn}, we shall show that it is correct
by checking that its variation leads to the classical equation of motion.
Suppose we vary the path $\ntau$ to $\ntau + \dta\ntau$.
$\dta\are$ is given by the area of the thin sliver
enclosed between the curves $\ntau$ and $\ntau +\dta\ntau$ (Fig.~3).
The part of this area due to the segments between $\tau$ and $\tau + \dta\tau$
equals
\bea
\Dta(\dta\are) &=& \Bigl[ \dta\ntau \times
                           [ \nhat(\tau + \Dta\tau) - \ntau] \Bigr]
                     \cdot\ntau \nnu\\
          &=& \left( \dta\ntau \times {d\nhat \by d\tau} \right)
                  \cdot \ntau \Dta\tau.
\eea
Adding up the contributions from all the segments, we find the total change
\beq
\dta\are = \int \dta\ntau \cdot
            \left({d\nhat \by d\tau} \times \ntau \right) \, d\tau.
\eeq
The variation of the second term in \eno{Sofn} is trivial, and
$\dta S$ can be written as an integral of the form
$\int\dta\ntau\cdot X$ where $X$ depends on $\nhat$ and $\Ecl$.
The extremal condition $\dta S = 0$ is thus
\beq
i J {d\nhat \by d\tau} \times \ntau + {\ptl \Ecl \by \ptl\nhat} = 0.
\eeq
Taking the cross product with $\nhat$, and using the
fact that $\nhat\cdot(d\nhat/d\tau) =0$, we get
\beq
iJ \dndt = - \left( \nhat \times \dedn \right). \lbl{eomJ2}
\eeq
This is exactly what we would get from \eno{eomJ} with $\bJ = J\nhat$
and the Wick rotation $t \to -i\tau$. In other words, it is
the imaginary time equation for Larmor precession in the effective
magnetic field $\ptl\Ecl/\ptl\nhat$.

The actual evaluation of $\are$ is often more easily done by using
Stokes's theorem to transform \eno{Asurf} to a line integral. Using
notation borrowed from electromagnetism, let us write
$d\Om = \bB\cdot\nhat \,ds$, where $\bB(\nhat) = \nhat$, and $ds$ is an area
element. The line integral is $\oint_{\crv}\bA \cdot d\nhat$, with
$\bB = \nabla\times\bA$.  Since $\bB = \nhat$, it is a monopole field,
which, as is known, cannot be represented in terms of a nonsingular vector
potential.  If this singularity is concentrated into a Dirac string at the
south pole, we can write
\beq
\ant = \oint_{\crv} \bigl[ 1 - \cos\tta(\phi) \bigr] \,d\phi,
   \lbl{Aline}
\eeq
where $\crv$ is viewed as parametrized by $\phi$. This formula is correct
as long as $\crv$ does not pass through the south pole, and provided one
increments or decrements $\phi$ by $2\pi$ every time one crosses
the date line.

\subsection{Application to \Fe8}
We now apply the above formalism to a discussion of the ground state
tunnel splitting in \Fe8 when $\bH\|\xhat$, further limiting ourselves to
finding the action, as this is sufficient to determine the field points
where the splitting is quenched.

The main problem is to calculate the action for an instanton. To do this,
however, we do not need to find the actual time dependence $\nhat(\tau)$,
since the energy is conserved along an instanton. To see this, note that
by \eno{eomJ2}
\bea
{d\Ecl \by d\tau} &=& \dedn\cdot\dndt \nnu\\
                  &=& {i \by J} \dedn \cdot \left(
                        \nhat \times \dedn \right) \nnu\\
                  &=& 0. \label{dEdt0}
\eea
Thus, using energy conservation,
we can find the instanton orbit, i.e., the curve traced out on
the unit sphere $\tta(\phi)$, without finding the actual time dependence.
Using \eno{Aline}, this is enough to find the action, as we can take
$\Ecl$ itself to be zero by adjusting the zero of energy appropriately.
One point to note is that since $\nhat_1$ and $\nhat_2$ are minima of
$\Ecl$, there can be no real curve connecting these points on which
$\Ecl$ is the same. The only way to find a solution to the instanton
equations of motion is to let $\nhat$ become complex. Correspondingly,
the area $\are$ must also be defined on the complexified unit sphere.

The above discussion is applicable to any tunnelling problem. What is
special about \Fe8 is the existence of {\it two\/} interfering
instantons.   If we choose the polar axis to be $\xhat$ (not $\zhat$),
and measure the azimuthal angle in the {\it yz\/} plane from $\yhat$,
then we can write the energy as
\beq
\Ecl(\tta,\phi) = k_1 J^2 (\cos\tta - \cos\tta_0)^2
                  + k_2 J^2 \sin^2\tta\sin^2\phi.
\eeq
We have defined $\cos\tta_0 = H/H_c$, with $H_c = 2k_1 J/g\mu_B$,
and added a constant to $\Ecl$ so that $\Ecl = 0$ along the instanton
as discussed above. Writing $\cos\tta_0 = u_0$, the solution
of this equation gives
\beq
\cos\tta = {u_0 +
              i \lam^{1/2} \sin\phi (1 - u_0^2 - \lam \sin^2\phi)^{1/2}
                          \by
             1 - \lam \sin^2\phi}. \label{inst}
\eeq
However, it is clear that in this equation, we may take $\phi$ to lie
in either of the two intervals $(0,\pi)$ and $(0,-\pi)$. Indeed, it
is obvious from symmetry that there are two instanton paths, which wind
about $\xhat$ in opposite directions (see Fig.~4), and \eno{inst}
describes both of them. If we denote the two paths by
A and B, then the imaginary parts of their actions ($=iJ\are$)
are necessarily unequal, since by
the interpretation of $\are$ as an area,
\beq
S_B - S_A = i J \times \Om,
\eeq
where $\Om$ [see \eno{Asurf}] is the area enclosed between A and B
(the peanut shape in Fig.~4). Using the line integral form
(\ref{Aline}), we obtain
\beq
\Om = \int_{-\pi}^{\pi} \left(1 - {u_0 \by 1 - \lam\sin^2\phi} \right) d\phi
       = 2\pi \left( 1 - {u_0 \by \sqrt{1-\lam}} \right).
\eeq
\begin{figure}
\centerline{\psfig{figure=\figdir/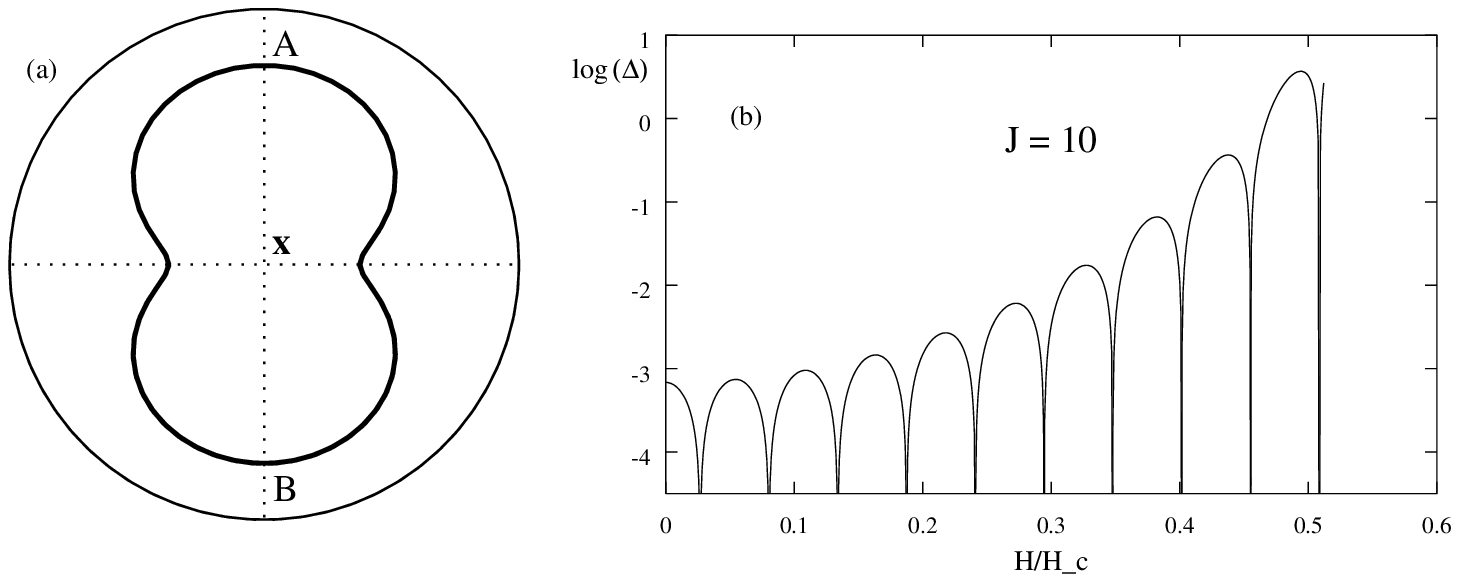,width=\linewidth}}
\caption{Interfering instanton trajectories, and numerically computed
$-10 \tofro +10$ tunnel splitting for a model for \Fe8 with $\bH\|\xhat$.}
\end{figure}

The real parts, on the other hand are equal and given by
\bea
S_R &=& {\rm Re}S_{A,B} = J \lam^{1/2} \int_0^{\pm \pi}
              {\sin\phi (1 - u_0^2 - \lam \sin^2\phi)^{1/2}
                          \by
             1 - \lam \sin^2\phi} d\phi, \nnu\\
    &=&  J \left[
           \ln \lp {\rta + \rtl \over \rta - \rtl} \rp
           - {u_0 \over \rtlb}
              \ln \lp {\rtd + u_0 \rtl \over \rtd - u_0 \rtl} \rp
            \right].  \label{Sr}
\eea

The prefactors $D$ are equal for both paths by symmetry, and so, adding
together the two contributions, we get
\beq
\Dta = 2D \exp(-S_R) \cos(J\Om/2), \label{Dtaans} 
\eeq
which vanishes whenever
\beq
{H \by H_c} = {\sqrt{1 - \lam} \by J}\Bigl[ J - n - \tshf \Bigr],
 \label{qpts}
\eeq
where $n=0, 1, \ldots, 2J - 1$.

In Fig.~4 we also show the results of an explicit numerical diagonalization
of the $21 \times 21$ Hamiltonian matrix for our model Hamiltonian for
\Fe8 showing clearly that the effect is genuine.
It should also be clear that the quenching phenomenon is a general
one that will occur as long as $\ham$ has the requisite symmetry, and
that the detailed form will only change the locations of the quenching
points. The minima in the tunnelling rate seen by Wernsdorfer and Sessoli
are spaced approximately 50\% further apart than implied by \eno{qpts}.
These differences are well understood in terms of higher order anisotropy
corrections to the model Hamiltonian (\ref{ham0}). These corrections do
not change the centers of gravity of the levels by very much, but they
affect the splittings significantly.

\section{Discrete phase integral approach to spin tunnelling}

It will not have escaped the reader that the Wernsdorfer and Sessoli
data (Fig.~2b) also show an oscillation in the tunnelling amplitude for
the $m=-10 \tofro +9$ and $-10 \tofro +8$ transitions. These oscillations
were not predicted in Ref.~\cite{agepl93}, and are nontrivial instances
of a diabolical point, or conical intersection \cite{bw84,hlh63}, a
degeneracy obtained at isolated points in a two dimensional parameter
space. In \Fe8, the parameters on which the Hamiltonian depends are
$H_x$ and $H_z$.

In retrospect, the new oscillations can also be understood in terms of
path integrals, but it is simpler to adopt a different point of view,
namely, the discrete phase integral (or WKB) method.
The basic idea of this method is to directly solve Schr\"odinger's
equation in the $J_z$ basis as a recursion relation or difference equation,
by making use of the similarity between difference and differential
equations, and exploiting WKB type ideas used to solve the latter.

To see what is meant, suppose that $\ket{\psi}$ is an eigenstate of
$\ham$ with energy $E$. With
$J_z \ket m = m\ket m$, $\tran{m}{\psi} = C_m$,
$\mel{m}{\ham}{m} = w_m$, and $\mel{m}{\ham}{m'} = t_{m,m'}$ ($m \ne m'$),
we have
\beq
\mathop{{\sum}'}_n t_{m,n} C_n + w_m C_m = E C_m, \label{Seq}
\eeq
where the prime on the sum indicates that the term $n=m$ is to
be omitted. This equation can be thought of as a tight binding model
for an electron in a one-dimensional lattice with sites labelled by $m$,
and slowly varying on-site
energies ($w_m$), nearest-neighbor ($t_{m,m\pm 1}$) hopping terms,
next-nearest-neighbor ($t_{m,m\pm 2}$) hopping terms, and so on.
Under certain conditions (which we shall see below) we can understand
the motion of the electrons in terms of wavepackets which are well
localized in both quasi-momentum and real space. This is of course
the approximation of semiclassical electron dynamics. As discussed
in standard texts on condensed matter physics, this approximations
works when the wavepacket can be assigned a real space width $\Dta m$
much less than the length scale over which the properties of
the electron band (in our case the $w_m$ and $t_{m,m+\al}$) vary,
and at the same time, a momentum space width $\Dta q$ much less than the 
reciprocal bandwidth $2\pi$. Since the two widths must be constrained
by the uncertainty principle $\Dta q \Dta m \sim 1$, the real space
width $\Dta m$ must be much greater than the lattice constant
(in our case 1). Formally, we need to be able to find smooth
continuum approximants $w(m)$ and $t_{\al}(m)$, such that
\bea
w(m) &=& w_m, \label{wcont} \\
t_{\al}(m) &=& (t_{m,m+\al} + t_{m,m-\al})/2, \quad \al=1,2,\ldots
   \label{tcont}
\eea
whenever $m$ is an eigenvalue of $J_z$, and that these quantities vary
slowly enough that
\beq
{dw \over dm} = O\left( {w(m) \over J } \rp, \quad
{dt_\al \over dm} = O\left( {t_{\al}(m) \over J } \rp.
\label{slow}
\eeq
These conditions are generally met when $J \gg 1$, which is also
the semiclassical limit for spin that we expect intuitively.

Given the conditions (\ref{slow}), the basic approximation, which readers will
recognize from the continuum case, is to write the
wavefunction as a linear combination of the quasiclassical forms
\beq
C_m \sim {1 \over \sqrt{v(m)}}\exp\lp i\int^m q(m') dm'\rp,
    \label{Cwkb}
\eeq
where $q(m)$ and $v(m)$ obey the equations
\bea
E &=& w(m) + 2\sum_{\al \ge 1} t_{\al}(m) \cos(\al q)
     \equiv \hsc(q,m), \label{hjeq} \\
v(m) &=& \ptl \hsc/\ptl q. 
     \label{vm}
\eea
Equations (\ref{hjeq}) and (\ref{vm}) are the lattice analogues of the
eikonal and transport equations. Equation (\ref{Cwkb}) represents
the first two terms in an expansion of $\log C_m$ in powers of
$1/J$.

Previous work with the DPI method \cite{dm67,sg75,pb78,pb93,vhs86}
has been limited to the case where the recursion relation has only three
terms, i.e.,  only nearest neighbor hopping is present.
Braun \cite{pb93} mentions many problems in many areas (nuclear and
atomic physics, to name just two) where the Schr\"odinger
equation turns into a three-term recursion relation in a suitable
basis. All the types of problems as in the continuum case
can then be treated---Bohr-Sommerfeld quantization, barrier
penetration, tunneling in symmetric double wells, etc. In addition, one
can also use the method to give asymptotic solutions for various
recursion relations of mathematical physics, such as those for the
Mathieu equation, Hermite polynomials, Bessel functions, and so on.
The interesting
features all arise from a single fact --- that the DPI approximation
breaks down at the so-called {\it turning points}. These are points
where $v(m)$ vanishes. One must relate the DPI solutions on opposite
sides of the turning point by connection formulas, and the solution
of all the diverse types of problems mentioned above depends on
judicious use of these formulas.

For the \Fe8 Hamiltonian (\ref{ham0}) with arbitrary $\bH$,  the
recursion relation involves five terms. 
The diagonal terms ($w_m$) arise
from the $J_z^2$ and $J_zH_z$ parts of $\ham$, the
$t_{m,m \pm 1}$ terms from the $J_xH_x$ and $J_yH_y$ parts, and the
$t_{m, m\pm 2}$ terms from the $J_x^2$ part. This seemingly
minor modification is responsible for all the beautiful spectral
features seen experimentally.

In the three term case, only $w(m)$ and $t_1(m)$ are present, so
$\hsc(q,m) = w(m) + 2t_1(m) \cos q$, and
$v(m) = -2\sin q(m) t_1(m)$. Hence, the turning points arise when $q=0$
or $q=\pi$, corresponding to the local $m$-dependent band edges. This
is completely analogous to the condition $V(x) = E$ that defines a
turning point for a massive particle moving in a potential $V(x)$, in
that a turning point is a limit of the classically allowed
range of motion.

In the five term case, by contrast, 
\beq
\hsc(q,m) = w(m) + 2t_1(m) \cos q + 2t_2(m) \cos(2q), \label{hftr}
\eeq
so that
\beq
v(m) = -2\sin q(m) \bigl(t_1(m) + 4 t_2(m) \cos q(m)\bigr).
     \label{vftr}
\eeq
In addition to $q=0$ and $q=\pi$, now the velocity can also vanish 
when $q = q_*(m)$, where
\beq
\cos q_*(m) = - t_1(m)/4t_2(m). \label{cosqstar}
\eeq
If we make use of the eikonal equation, we see that
a turning point arises whenever
\beq
E = U_0(m),\ U_{\pi}(m),\ {\rm or}\ U_*(m),
   \label{Econd}
\eeq
where,
\bea
U_0(m) &=& \hsc(0,m) = w(m) + 2t_1(m) + 2t_2(m), \label{U0} \\
U_{\pi}(m) &=& \hsc(\pi, m) = w(m) - 2t_1(m) + 2t_2(m), \label{Upi} \\
U_*(m) &=& \hsc(q_*,m) = w(m) - 2t_2(m) - {t_1^2(m) \over 4 t_2(m)}.
           \label{Ustar}
\eea
We call these curves {\it critical\/} curves; collectively they play
the same role as the potential energy $V(x)$ in the continuum phase
integral method in determinimg the turning points.

The new turning point $q_*(m)$ and the corresponding critical curve
$U_*(m)$ have no analogue in the continuum case. This turning point need
not be at the limit of the classically allowed energies for fixed $m$,
and may lie strictly inside the band, or in the forbidden region outside
it. We show the general form of the critical curves for \Fe8 when $\bH$
has both {\it x\/} and {\it z\/} components in Fig.~5. The point
$m'_c$ is of this new type, and here it may be said to lie ``under the
barrier."
\begin{figure}
\centerline{\psfig{figure=\figdir/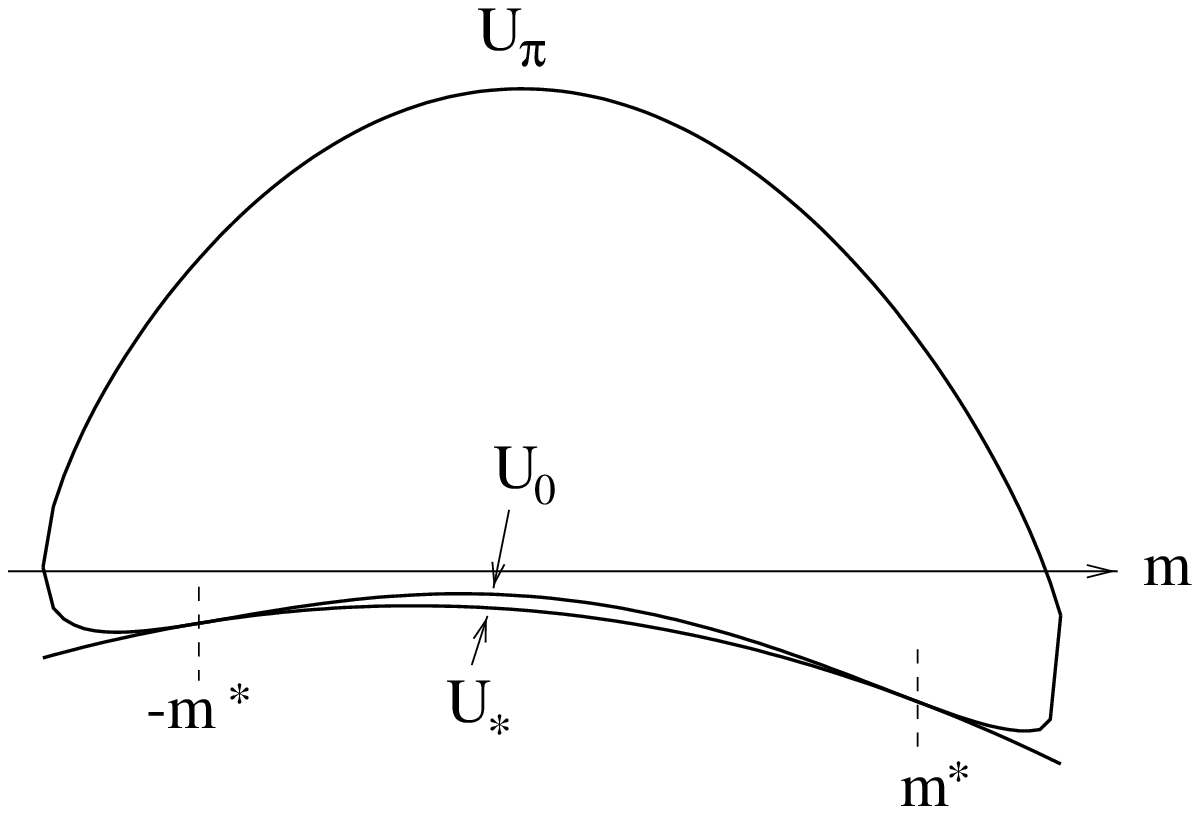,height=5cm,%
width=0.46\linewidth}
\psfig{figure=\figdir/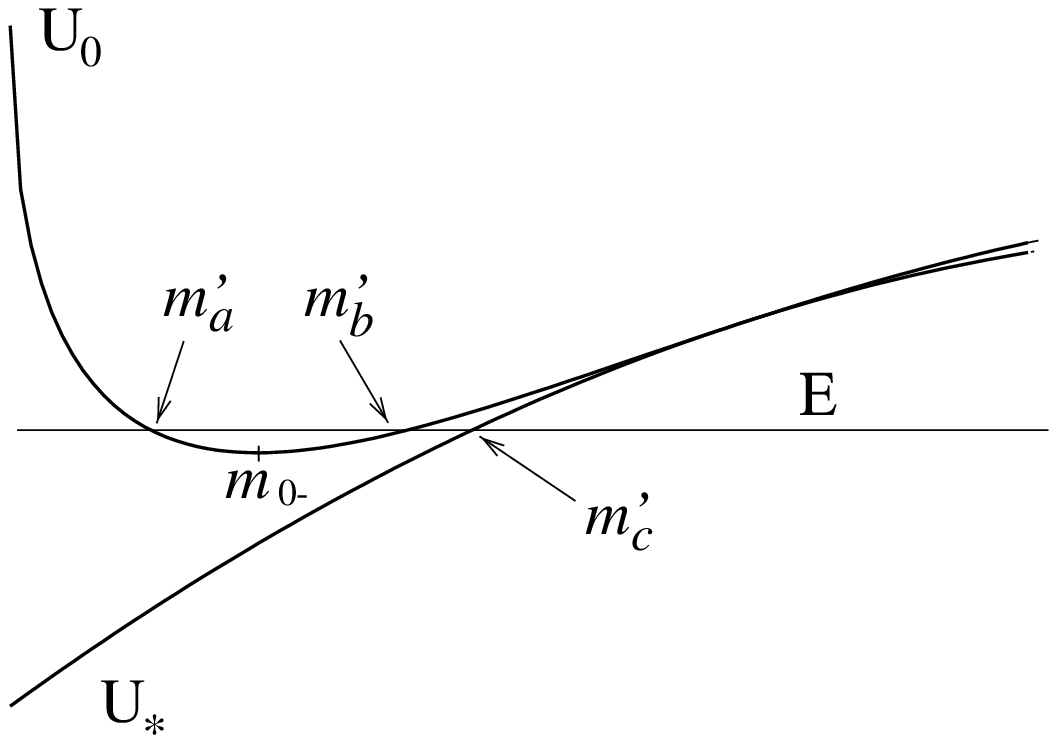,height=5cm,width=0.46\linewidth}}
\caption{Critical energy curves for the \Fe8 Hamiltonian when
$\bH$ has both $x$ and $z$ components.  In the right hand
figure, the region near the left minimum of $U_0$ is magnified,
showing the various turning points.}
\end{figure}

From this point on, the analysis of eigenfunctions and eigenvalues
is conceptually straightforward, although quite lengthy, and we
refer readers to already existing papers for the details
\cite{agprl99,vf00}. The basic idea is simple. For
an energy $E$ such as that shown in Fig.~5b, the region
$m'_a < m < m'_b$ is classically allowed, along with its counterpart
for the right hand well. To find the eigenfunctions, one starts
with a solution that is exponentially decaying to the left for
$m < m'_a$, continues it through the turning points $m'_a$, $m'_b$, and
$m'_c$ by the use of connection formulas at each point, and matches
it with a similar solution obtained by starting from the right. In general
the wavefunctions will agree in the central region near $m=0$ only
if the energy $E$ is properly chosen---this gives the eigenvalue
condition. Nevertheless, as already stated, the calculation has several
novel aspects. For example, refering to Fig.~5b, one should
clearly have ${\rm Im}\,q(m) < 0$ in the region $m < m'_a$. In conventional
WKB, there would be one solution satisfying this demand, but now there are
{\it two}. As another example, the two solutions in the region
$m'_b < m < m'_c$ which are both decaying exponentially to the right,
turn into decaying solutions with an oscillatory envelope
as the point $m'_c$ is crossed. It is this
oscillatory exponential solution (along with the concommitant fact that
there must be two such solutions when they exist) that allows the
tunnel splitting to vanish if $H_x$ and $H_z$ are chosen properly.
Cutting short an already long tale, we give the results for the location
of these points. The point where the $\ell'$th level in
the negative $J_z$ well (with $\ell'=0$ being the lowest level) and the
$\ell''$th level in the positive one are degenerate, is at $H_y = 0$, and
\bea
{H_z(\ell',\ell'') \over H_c}
    &=& {\sla (\ell'' - \ell') \over 2 J}
                 \label{plhz} \\
{H_x(\ell',\ell'') \over H_c}
   &=& {\sqrt{1-\lam} \over J}
        \left[ J - n - \tshf (\ell' + \ell'' + 1) \right],
             \label{plhx}
\eea
with $n = 0, 1, \ldots, 2J - (\ell' + \ell'' + 1)$. Here,
$\lam = k_2/k_1$, and $H_c = 2 k_1 J/g\mu_B$. Exactly as seen in the
experiments, the $-10\tofro 9$ points are shifted by half a period
with respect to the $-10\tofro 10$ points, the $-10 \tofro 8$ points
are shifted by another half-period, and so on. Further, many of the
degeneracies occur simultaneously, i.e., at the same values of
$H_x$ and $H_z$. This latter feature has only been tested indirectly
in the experiments so far.

We conclude with an amusing fact about Eqs.~ (\ref{plhz}) and (\ref{plhx}).
These are the leading $1/J$ results from a semiclassical analysis.
Nevertheless, for the model (\ref{ham0}), they are exact as written!
This has been demonstrated
in Ref.~\cite{kg00}, and while the exactitude is spoilt by the higher
order anisotropies present in real \Fe8, the exact results may pave the
way for a quantitative treatment of this and other perturbations.

\acknowledgments
This article is based on a paper presented at the Winter Institute on the
Foundations of Quantum Theory, S.~N. Bose Centre, Calcutta, in January
2000. I am grateful to the organizers of the conference for the
invitation to the workshop, and to the staff and members of the Bose
Institute for their cordial hospitality. The research reported herein is
supported by the U.~S. National Science Foundation, thorugh grant number
DMR-9616749.

\end{document}